# Constructing the Critical Curve for the Two-Layer Potts Model Using Cellular Automata

Yazdan Asgari[1] , Mehrdad Ghaemi[2]

[1] The Center for Complex Systems Research, K.N.T. University of Technology, Tehran, Iran
yazdan1130@hotmail.com
[2] Department of chemistry, Teacher Training University, Tehran, Iran
ghaemi@tmu.ac.ir

**Abstract**

The critical points of the 3-states two-layer Potts model on square lattice for different interlayer couplings ($K_x$=$K_y$≠$K_z$) are calculated with high precision using probabilistic cellular automata with Glauber algorithm, where $K_x$ and $K_y$ are the nearest-neighbor interactions within each layer in the $x$ and $y$ directions, respectively and $K_z$ is the interlayer coupling. The obtained results are 0.726, 0.807, 0.928, 0.987 and 1.00 for the cases $K_z/K_x$ = 1, 0.5, 0.1, 0.01 and 0.001, respectively. Then, the critical curve has been constructed for this model.

## Introduction

Since it was known that the Ising model and its variants can be used to describe many critical phenomena, there have been a lot of attempts to solve various Ising models and to obtain critical points and critical exponents. Since the exact solution of Ising models exists only for one and two dimensional models[1,2], and there is no such solution for others, the simulation methods or other numerical methods may be used to get critical data[3-7].

There are different numerical methods which are limited to lattices with finite sizes. Their critical points are calculated using extrapolation approach. One of these methods is using the transfer matrix and decreasing the matrix size. For instance, in recent paper[8], it was shown that one could calculate the critical point by decreasing the matrix size for the two-layer three state Potts model. Such calculations are limited to the square two-layer lattice with width 5 cells in each layer and the critical point is obtained by extrapolation approach. Numerical methods are time consuming and advanced mathematics is required.

Simulation methods like Monte Carlo[9] are generally simpler than numerical methods and also are limited to lattice with finite size. But the Cellular Automata (CA) are one of the simulation methods that seems to be a good candidate to calculate the critical data. In order to make calculations faster and more precise, the CA can be parallelized. In the CA, in spite of the Monte Carlo method, all cells are updated simultaneously. Although the numerical methods are based on the calculation of a specific quantity, all the desire information may be calculated in each run in the CA.

So, using the CA led to perform many works describing Ising models and a great number of papers and excellent reviews were published[10-16]. Because of the challenge to discover faster algorithm for its simulation, different approaches have been used. For example, the Q2R automaton and its limitations was suggested and has been studied extensively[16-21]. It was so fast, because no random numbers must be generated at each step. But in the probabilistic CA for Ising models like Metropolis algorithm[22], generation of random number makes it slower, even though it is more realistic for description of the Ising model. Most of the works that have been done until now are for qualitative descriptions or for introducing fast methods for solution of various Ising models. The aim of our approach is to show that the CA increases the precision of the calculations.

In recent work, Asgari et al.[23] showed that the CA could be used to obtain critical points of the two-layer Ising and the two-layer 3-state Potts models with a high precision for the isotropic and symmetric case ($K_x=K_y=K_z=K$), where $K_x$ and $K_y$ are the nearest-neighbor interactions within each layer in $x$ and $y$ directions, respectively, and $K_z$ is the interlayer coupling.

In the present work, we have extended our approach to an asymmetric case for the two-layer Potts model ($K_x=K_y \neq K_z$) in a square lattice with Glauber algorithm[24]. We have shown that obtained values are in good agreement with our expectation. Finally, we have fitted the obtained values of the asymmetric two-layer Potts model into a polynomial of at least order of three and constructed the critical curve for this model.

In section one; we have introduced a 3-states two-layer Potts model. Then in section two, we have described our approach in detail for obtaining the critical points for different interlayer couplings and finally, the critical curve for this model is constructed.

## 1 Two-layer Potts Model

Although we do not know the exact solution of the Potts model for any two-layer at present time, a large amount of numerical information has been accumulated for the critical properties of the various Potts models[25]. For further information, see the excellent review written by Wu[5] or references given by him.

Consider a two-layer square lattice with the periodic boundary condition, each layer with $p$ rows and $r$ columns. Each layer has then $r \times p$ sites and the number of the sites in the lattice is $2 \times r \times p = N$. We consider the next nearest neighbor interactions as well, so the number of neighbors for each site is 5. For any site we define a spin variable $\sigma^{1(2)}(i,j) = 0,\pm 1$ so that $i = 1,...r$ and $j = 1,...,p$. The configurational energy of a standard 3-state Potts model is given as[23,5] ,

$$\frac{E(\sigma)}{kT} = \sum_{i=1}^{r,*}\sum_{j=1}^{p,*}\sum_{n=1}^{2} -\{K_x \delta_{\sigma^n(i,j),\sigma^n(i+1,j)} + K_y \delta_{\sigma^n(i,j),\sigma^n(i,j+1)} + K_z \delta_{\sigma^1(i,j),\sigma^2(i,j)}\} \tag{1}$$

where

$$\delta_{i,j} = 1 \text{ for } i = j$$

$$\delta_{i,j} = 0 \text{ for } i \neq j \tag{2}$$

and * indicates the periodic boundary condition and $K_x$ and $K_y$ are the nearest-neighbor interactions within each layer in $x$ and $y$ directions, respectively, and $K_z$ is the interlayer coupling. Therefore, the configurational energy per spin is

$$e = \frac{E(\sigma)}{kTN} \tag{3}$$

The average magnetization of the lattice for this model can be defined as[9]

$$\langle M \rangle = \left\langle \sum_{i=1}^{r,*}\sum_{j=1}^{p,*}\sum_{n=1}^{2} \sigma^n(i,j) \right\rangle \tag{4}$$

and the average magnetization per spin is

$$\langle m \rangle = \frac{\langle M \rangle}{N} \tag{5}$$

The magnetic susceptibility per spin ( $\chi$ ) and specific heat per spin ($C$) is defined as[9]

$$\frac{\partial < M >}{\partial \beta} = \beta(< M^2 > - < M >^2) \tag{6}$$

$$\chi = \frac{\beta}{N}\left(\langle M^2 \rangle - \langle M \rangle^2\right) = \beta N\left(\langle m^2 \rangle - \langle m \rangle^2\right) \tag{7}$$

$$C = \frac{k\beta^2}{N}\left(\langle E^2 \rangle - \langle E \rangle^2\right) = k\beta^2 N\left(\langle e^2 \rangle - \langle e \rangle^2\right) \tag{8}$$

where $\beta = \frac{1}{kT}$.

## 2 Method

The algorithm for the automaton process is given in appendix A. For quantitative computation of the critical temperature of a two-layer 3-state Potts model, we considered the isotropic ferromagnetic and asymmetric case which $K_x=K_y \neq K_z \geq 0$. We have used a two-layer square lattice that each layer has $1500 \times 1500$ sites and to reduce the finite size effects the periodic boundary condition is used. Each site can have a value of +1, -1 or zero. We used the Glauber method with checkerboard approach to update the sites. Namely, each layer is like a checkered surfaces and at first, the updating is done for the white parts of the first layer. Then the black ones are updated. After which, this approach is done for the second layer. The updating of +1 spins is based on the probabilistic rules. The probability that spin of one site will be +1 ( $p_i^+$ ) is given by

$$p_i^+ = \frac{e^{-\beta E_i^+}}{e^{-\beta E_i^+} + e^{-\beta E_i^-} + e^{-\beta E_i^0}} \tag{9}$$

Hence, probability that a given spin to be -1 ( $p_i^-$ ) is

$$p_i^- = \frac{e^{-\beta E_i^-}}{e^{-\beta E_i^+} + e^{-\beta E_i^-} + e^{-\beta E_i^0}} \tag{10}$$

and for the zero state we have,

$$p_i^0 = 1 - (p_i^+ + p_i^-) \tag{11}$$

where

$$E_i^{\pm,0} = -K_x\{\delta_{\sigma^n(i,j),\sigma^n(i+1,j)} + \delta_{\sigma^n(i,j),\sigma^n(i-1,j)}\} - K_y\{\delta_{\sigma^n(i,j),\sigma^n(i,j+1)} + \delta_{\sigma^n(i,j),\sigma^n(i,j-1)}\} - K_z\{\delta_{\sigma^n(i,j),\sigma^{n'}(i,j)}\} \tag{12}$$

and

$$\sigma^n(i,j) = +1 \text{ for } E_i^+$$

$$\sigma^n(i,j) = -1 \text{ for } E_i^-$$

$$\sigma^n(i,j) = 0 \text{ for } E_i^0 \tag{13}$$

It should be mentioned that in our approach, first we construct the probability matrix according to Eqs. (9-11) for different states of a cell in such a way that for each state it is sufficient to refer to the probability matrix and use the proper value of the probability. This leads to prevent similar calculations.

As we showed in the previous work[23], the critical point for such models could be obtained from different approaches. In another word, there are several ways to obtain the critical point. Here, we review these approaches as follow. When we start the CA with the homogeneous initial state (namely, all sites have spin up or +1), before the critical point ($K_c$), the magnetization per spin (*m*) will decay rapidly to zero and fluctuate around that point. After the critical point, *m* will approach to the nonzero point and fluctuate around it and with increasing of *K*, the magnetization per spin will go toward its initial state (i.e. $m = +1$). But at the critical point, *m* will decay very slowly to zero with a great fluctuation. For each value of *K*, the time that *m* reaches to a special value and starts to fluctuate, is called the relaxation time (τ). On the other hand, the relaxation time is the time that system is thermalized. The value of τ can be obtained from the graph of *m* versus *t*. So one can see from this graph that the relaxation time increases before the critical point and is maximum at $K_c$, but after the critical point, τ decreases. So, in the critical point, the system last for long time to stabilize. Hence, the critical point could be obtained from the graph of τ versus $K$ (Fig. 2).

Another way to get the critical point is the usage of the thermodynamic quantities after thermalization of the lattice. In another word, first we let the system to reach to a stable state after some time step (t = τ). Next we let the system to be updated to the end of the automata (t = 100000). For example, to calculate the average value of magnetization per spin ($<m>$), one should add all of values for $m$ from the relaxation time to the end of the automata (or end of the time step) and divide the result to the numbers of steps. By drawing the graph of $<m>$ versus $K$, we could get $K_c$. In this graph, fore $K<K_c$, the value of $<m>$ lies around zero. But it becomes nonzero at $K=K_c$, after which, its value increases gradually. For calculation of the susceptibility per spin χ (eq. 7), for each $K$, first we calculated the value of $(m-<m>)^2$ in each time step. Then these values are averaged by the same method explained above. Also the calculation of the specific heat $C$ (eq. 8), may be done by a similar way. The graphs of χ versus $K$ and $C$ versus $K$, are another approach to obtain the critical point. The maximum of such graphs gives the critical point.

The result of such calculations are shown in figures 1-5 for the simplest case of the two-layer 3-state Potts model when $K_x=K_y=K_z=K\geq 0$. The obtained value for the critical point is 0.726 for this case.

In next step, we have extended the approach for the two-layer Potts model with different interlayer coupling coefficients ($K_z$) and obtained the critical points for those asymmetric states. The calculated values are shown in table1. Then, we have fitted the obtained results into a polynomial. As shown in Fig. 5, the calculated critical points fitted into a polynomial of at least order of three in terms of ξ as,

$$K_c(\xi) = c_0 + c_1\xi + c_2\xi^2 + c_3\xi^3 \tag{14}$$

where $\xi = K_z / K_x$ and the universal coefficients are: $c_0$ = 0.9981, $c_1$ = -0.8182, $c_2$ = 1.198, $c_3$ = -0.6518.

Since there are only five obtained values for the critical point, they could be fitted into a polynomial of order of maximum three.

Finally, it should be mentioned that the most important part of the calculation is the computation of the average quantities especially near the critical point. As we mentioned before the starting point for averaging is specified from the graph of $m$ versus $t$. As shown in figure 1, near the critical point, the fluctuation is very high and one should be careful to identify the equilibrium state to start the averaging. There are some techniques by which the precision for equilibrium determination increases. One way is to increase the number of lattice size. Such a way, causes the system to have a less fluctuation and so, determination of the equilibrium state is easier. Also, the number of time steps should be high enough to determine the starting point for averaging.

But it is clear that the increasing the number of time steps and lattice size, lead to decrease of the rate of the program. To decrease the computational time, we tabulated the probabilities obtained from eqs. 9-11. In another word, the calculated probabilities are tabulated. So, when the program is running in each update, it is sufficient to refer to such a table and find the desire values for different probabilities. Another way to increase the program rate is the method of parallel processing on cluster computers for the case of a large lattice size.

## 3 Conclusion

Unlike the other numerical approaches given in the literatures, the advantage of the calculation of the critical point using the probabilistic CA is that it is possible to get the fourth and more digits after the decimal point with a high precision.

We have obtained the critical curve as a third-order polynomial of $\xi$ for a 3-states two-layer Potts model. The importance of the third order polynomial is due to the fact that, the experimental data could be easily be fitted into a polynomial, from which a unique value is obtained for any physical property. One may extend such calculations to other lattice such as triangular, hexagonal, and also other models like multi-states two-layer Potts model, 3-D Ising model, asymmetric two-layer models,… and use the results for modeling physical systems.

## Acknowledgment

We acknowledge Prof. G. A. Parsafar for his useful comment.

## Appendix A

Here, the algorithm of our approach for the two-layer three state Potts model is described;

Definition of the first lattice

Definition of the second lattice

Definition of different interlayer couplings ($K_x$, $K_y$, $K_z$)

Considering a up-spin (+1) for each cell

Construction the probability matrix according to Eqs. (9-11)

Beginning of the automata

- Choose all cells one by one
- By knowing the state of the nearest neighbors of a given cell, we may use the probability matrix to calculate the probability for each state of the cell
- Generation a random number
- By comparing the random number with the probability, the cell state may be +1, -1, or 0.

(Note that the above procedure must be carried out for all the cells of the two layers with a checkerboard approach described in section two.)

Calculation of the desired quantities according to Eqs. (3-8)

End of the automata

**Table caption**

Table1

The critical points for different interlayer coupling coefficients for the 3-states two-layer Potts model

Table 1

| $\xi = \frac{K_z}{K_x}$ | $K_C$ |
|---|---|
| 1 | 0.726 |
| 0.5 | 0.807 |
| 0.1 | 0.928 |
| 0.01 | 0.987 |
| 0.001 | 1.00 |

**Figure Captions**

Figure 1

Magnetization versus time for 3 states. a: $K$=0.721 ($K<K_c$). b: $K$=0.726($K=K_c$). c: $K$=0.729 ($K>K_c$). (each layer has $1500\times1500$ sites, start from homogeneous initial state "all +1", time steps = 100000)

Figure 2

Relaxation time ($\tau$) versus coupling coefficients ($K$). (calculated data are the results of the lattice with $1500\times1500$ sites in each layer, start from homogeneous initial state with all of the spins up, time steps = 100000)

Figure 3

$<m>$ versus coupling coefficients ($K$). (calculated data are the results of the lattice with $1500\times1500$ sites in each layer, start from homogeneous initial state with all of the spins up, time steps = 100000)

Figure 4

Magnetization susceptibility per spin ($\chi$) versus $K$. (calculated data are the results of the lattice with $1500\times1500$ sites in each layer, start from homogeneous initial state with all of the spins up, time steps = 100000)

Figure 5

Specific Heat per spin ($C$) versus $K$. (calculated data are the results of the lattice with $1500\times1500$ sites in each layer, start from homogeneous initial state with all of the spins up, time steps = 100000)

Figure 6

The values of $K_c$ versus $K_z/K_x$ for the two-layer Potts model which is fitted into a polynomial of order of three (dotted line).

Figure 1

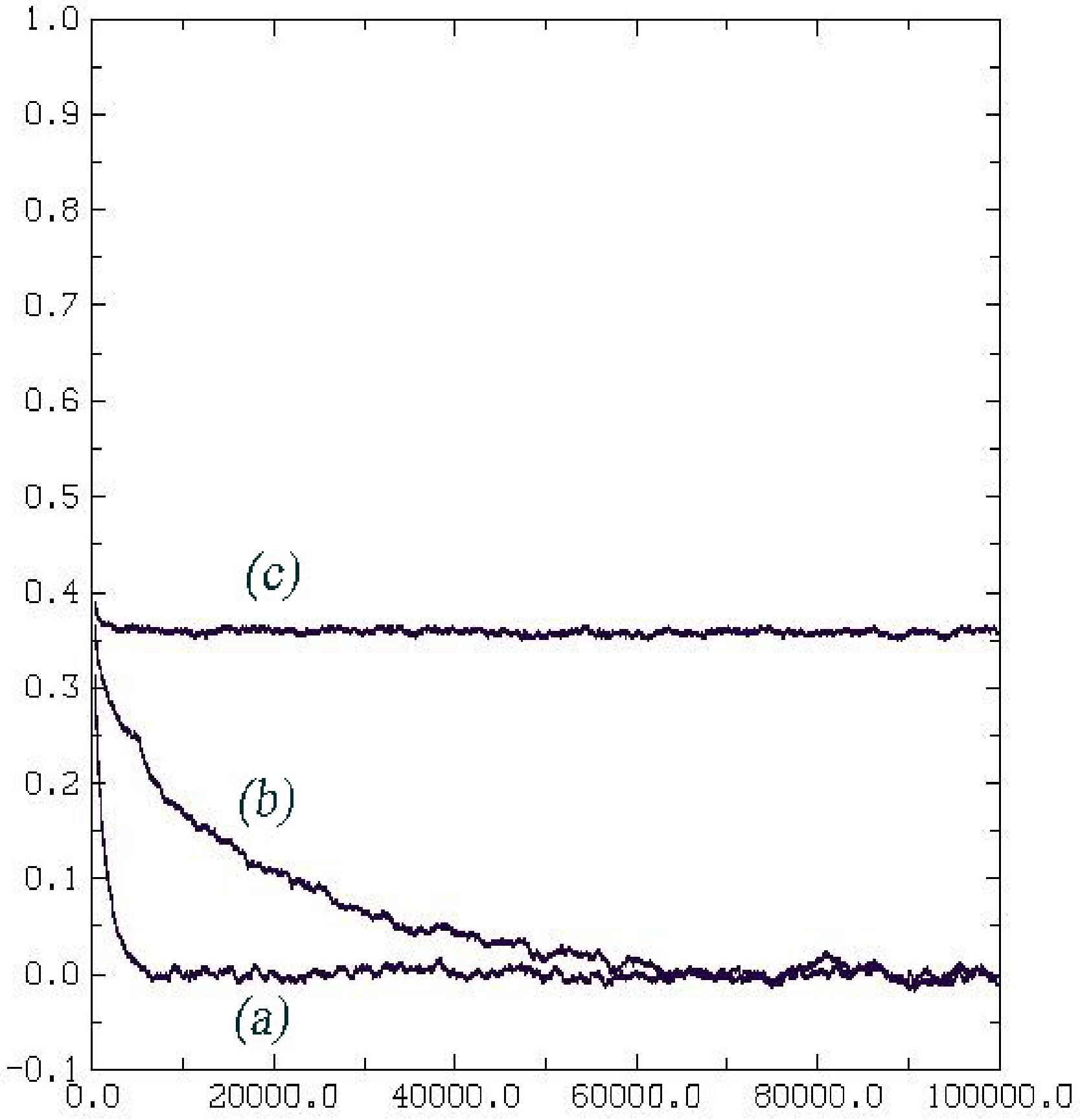

Figure 2

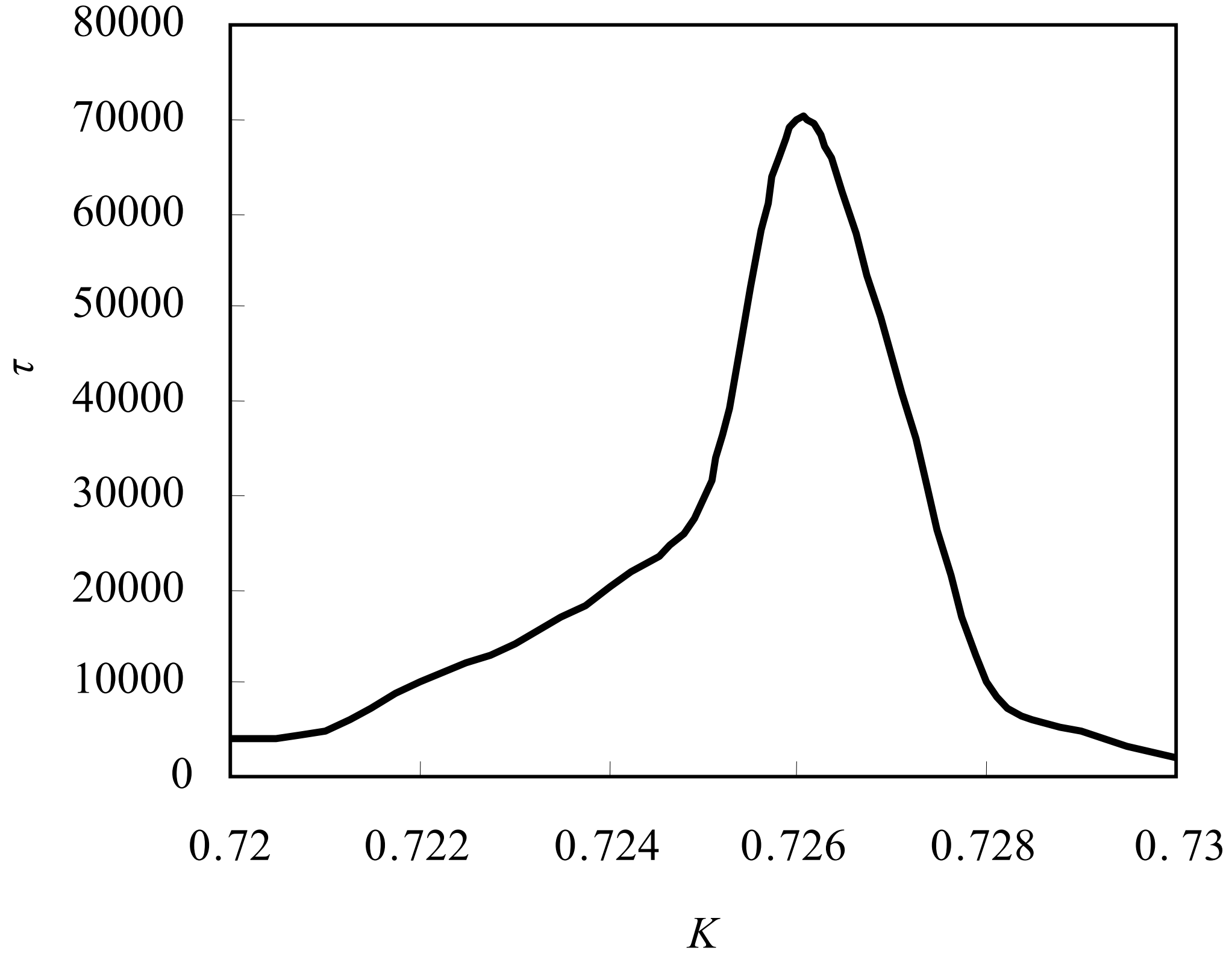

Figure 3

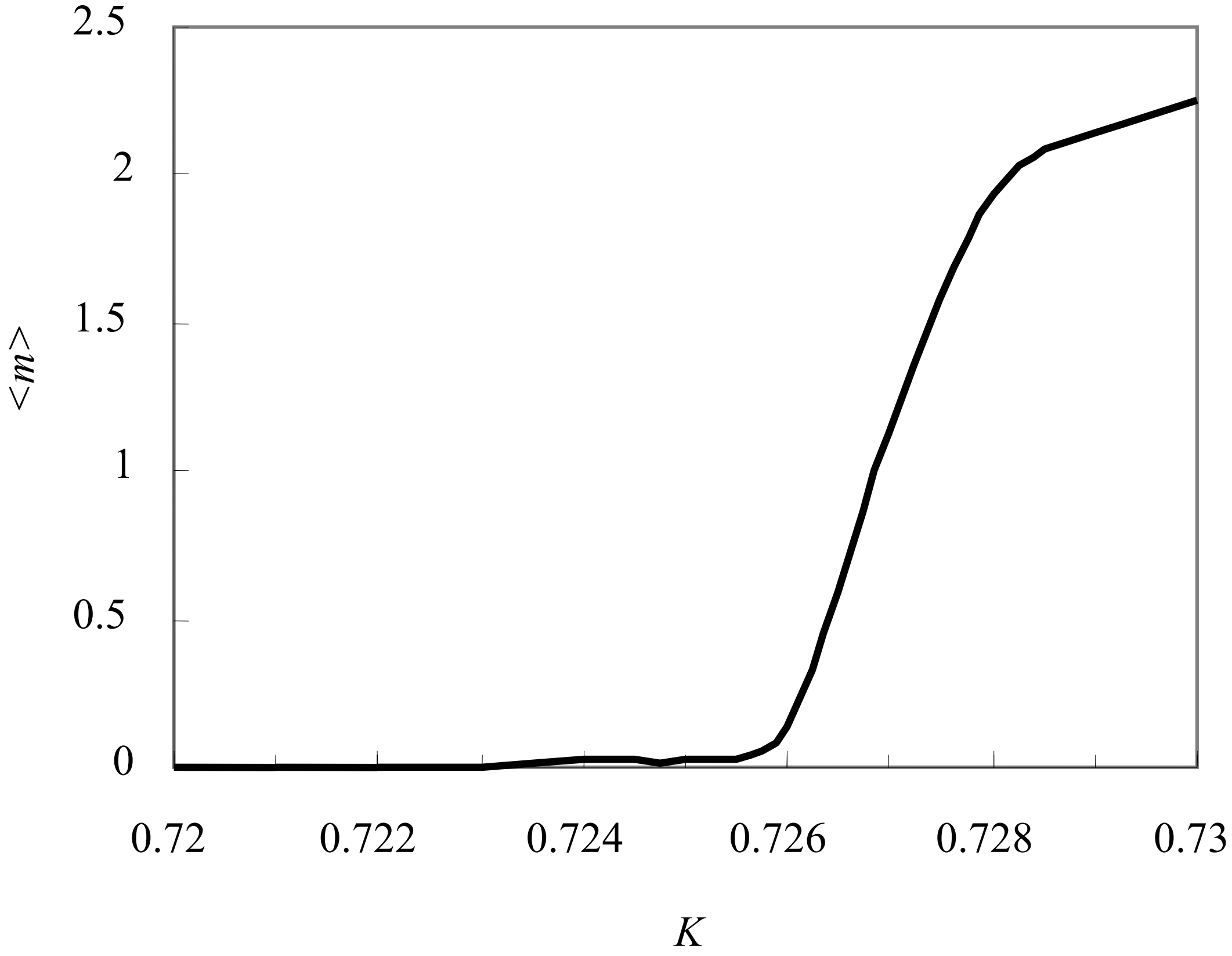

Figure 4

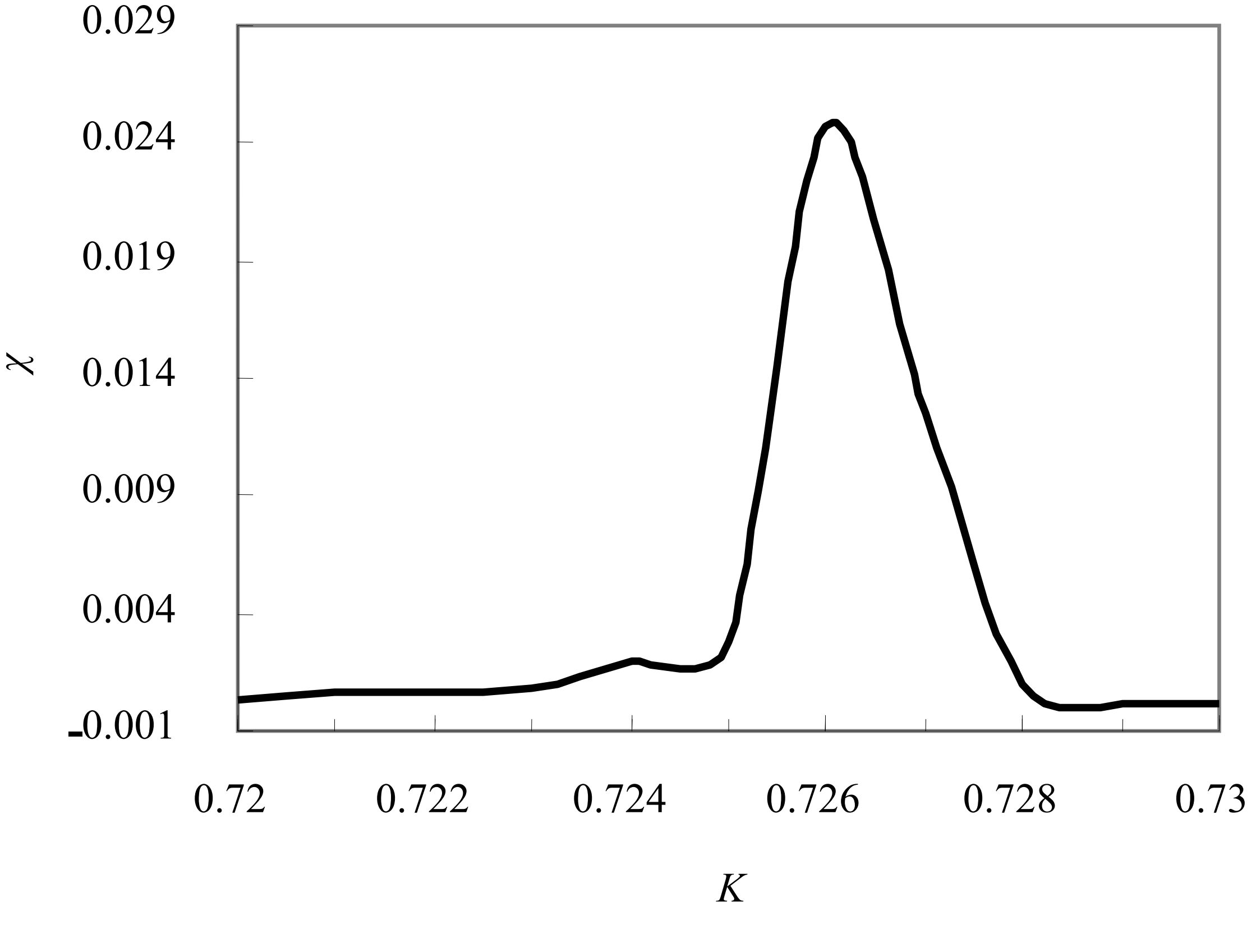

Figure 5

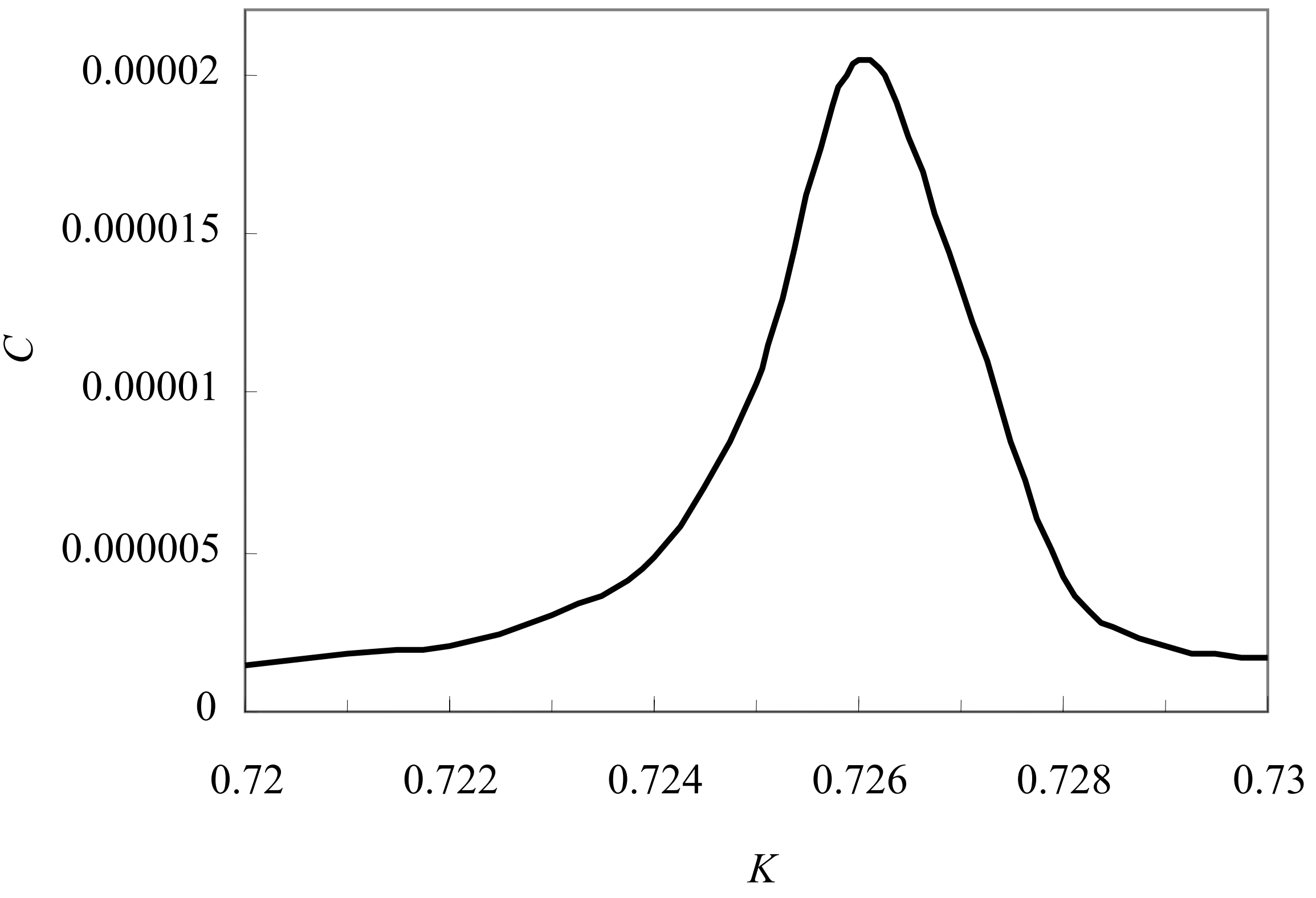

Figure 6

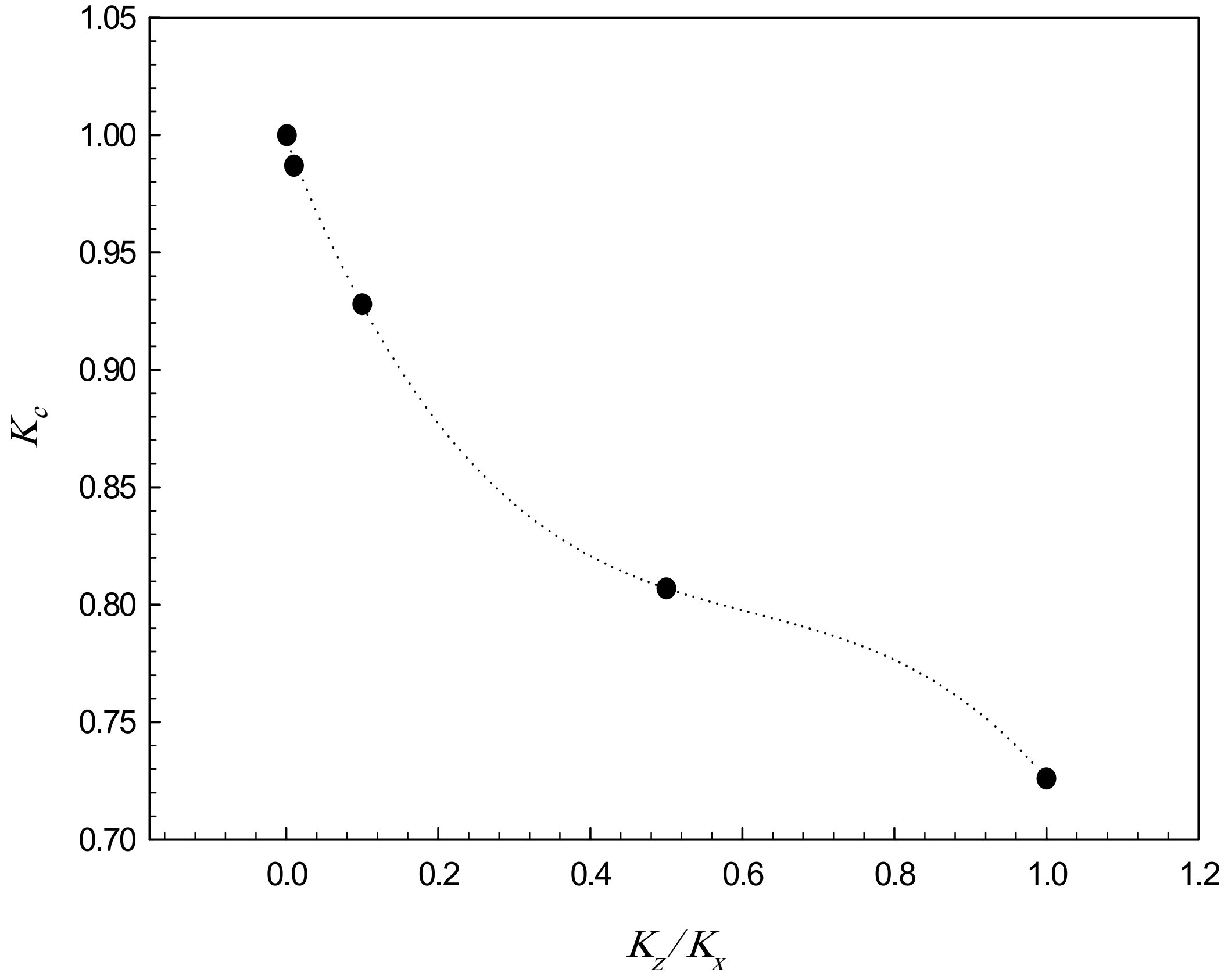